\documentclass[11pt]{article}
\usepackage{moriond}
\input{psfig.tex}
\usepackage{epsf}
\usepackage[dvips]{graphicx}

\newcommand{\beq}{\begin{equation}}
\newcommand{\eeq}{\end{equation}}
\newcommand{\beqa}{\begin{eqnarray}}
\newcommand{\eeqa}{\end{eqnarray}}

\newcommand{\laem}{\begin{array}{c} < \vspace{-0.5em} \\ {\scriptstyle \sim}
\end{array}}
\newcommand{\gaem}{\begin{array}{c} > \vspace{-0.5em} \\ {\scriptstyle \sim}
\end{array}}

\def\slashchar#1{\setbox0=\hbox{$#1$}           % set a box for #1
   \dimen0=\wd0                                 % and get its size
   \setbox1=\hbox{/} \dimen1=\wd1               % get size of /
   \ifdim\dimen0>\dimen1                        % #1 is bigger
      \rlap{\hbox to \dimen0{\hfil/\hfil}}      % so center / in box
      #1                                        % and print #1
   \else                                        % / is bigger
      \rlap{\hbox to \dimen1{\hfil$#1$\hfil}}   % so center #1
      /                                         % and print /
   \fi}                                         %

\def\be{\begin{equation}}
\def\ee{\end{equation}}
\def\bea{\begin{eqnarray}}
\def\eea{\end{eqnarray}}

\begin{document}
\vspace*{4cm}

\title{Limits on the Mass of a Composite Higgs Boson\,\footnote{For a more complete
discussion, see ref. 1 and references therein.}
}

\author{R. Sekhar Chivukula}

\address{Department of Physics, Boston University, 590 Commonwealth Ave.,\\
Boston, MA 02215 USA}

\maketitle\abstracts{We discuss the bound on the mass of the Higgs
  boson arising from precision electroweak measurements in the context of the
  triviality of the scalar Higgs model. We show that, including
  possible effects from the underlying nontrivial dynamics, a Higgs
  boson mass of up to 500 GeV is consistent with current
  data.}

Current results from the LEP Electroweak Working
Group\,\cite{moriondewwg,lepewwg} favor a Higgs boson mass that is
relatively light. The ``best-fit'' value for the Higgs mass is
somewhat less than experimental lower bound of 107.7 GeV reported at
this conference.\cite{moriondhiggs} The 68\% and 95\% CL upper bounds
from precision measurements, in the context of the standard model, are
127 and 188 GeV respectively.  It is possible that, as these data
suggest, the Higgs boson lies around the corner and will be discovered
at relatively low masses. On the other hand, it is important to
consider alternatives and to understand the wider class of models
consistent with precision electroweak tests.  In this talk, I will
show that even minor modifications to the standard electroweak theory
allow for a substantially heavier Higgs boson.

\section{The Triviality of the Standard Higgs Model}

\begin{figure}[htb]
\begin{center}
\begin{minipage}[t]{7.5cm}
\includegraphics*[bb=100 95 640 515,width=7.5cm]{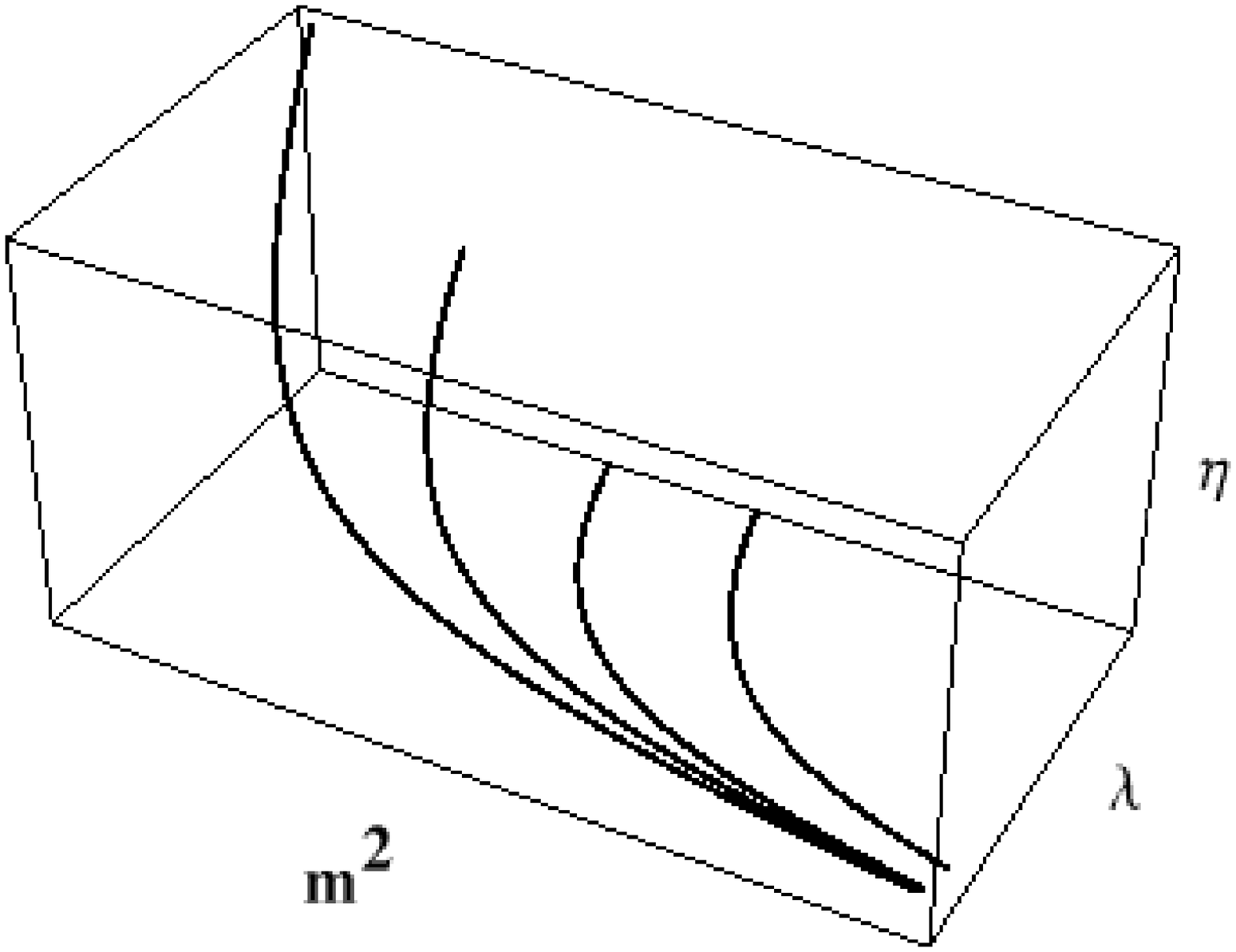}
\caption {Graphical representation of Wilson RG flow of 
  $(m^2(\Lambda),\lambda(\Lambda),\eta(\Lambda))$. As we scale to low
  energies, $m^2 \to \infty$, $\lambda \to 0$, and $\eta \to 0$.}
\label{3drg}
\end{minipage}
\hspace{2mm}
\begin{minipage}[t]{7.5cm}
\includegraphics*[width=7.5cm]{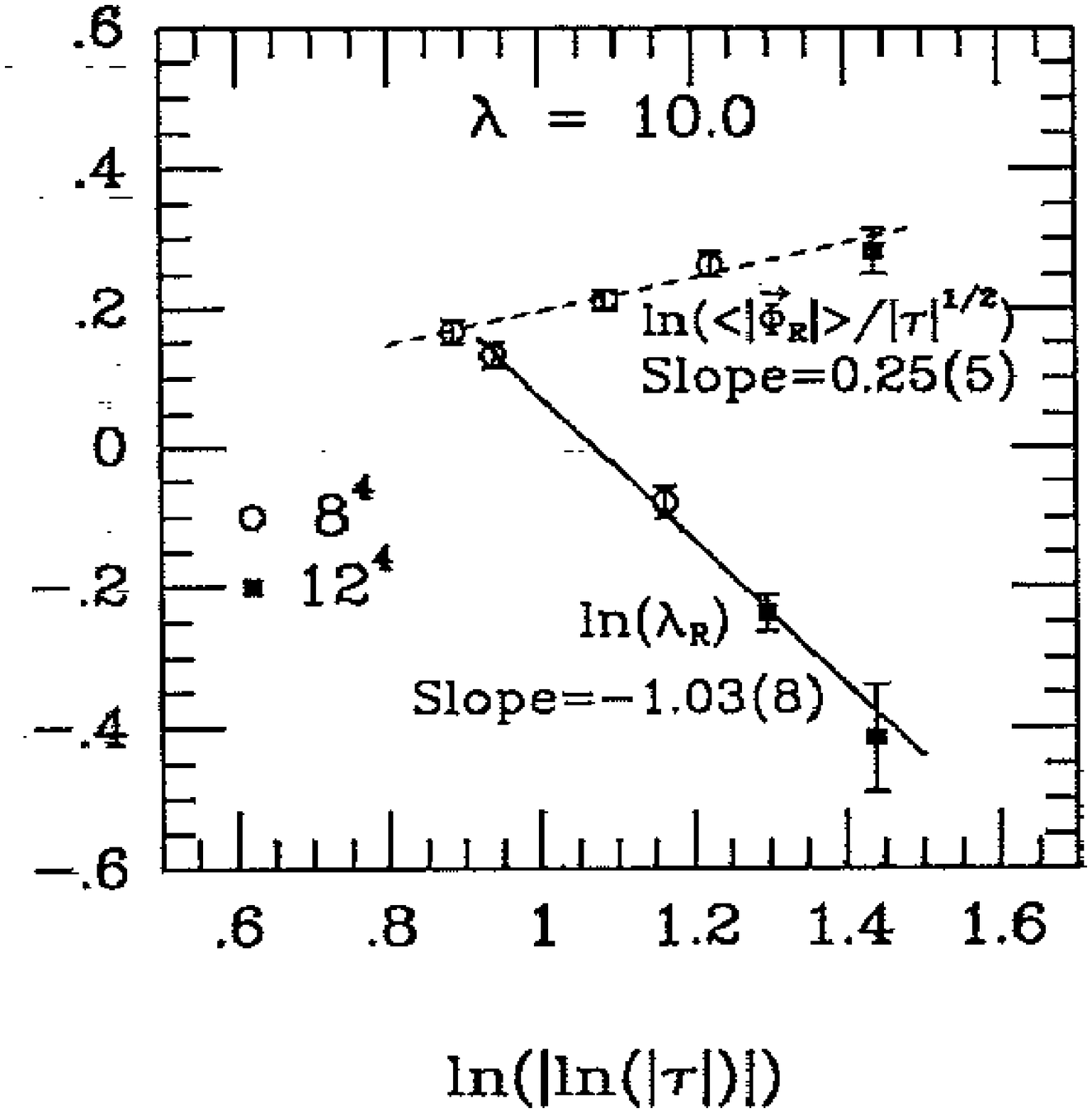}
\caption{Results of a nonperturbative lattice monte carlo 
  study\,\protect\cite{kuti} of the scalar sector of the standard model
  with bare coupling $\lambda=10$. The approximate slope of -1 for the
  renormalized coupling, $\lambda_R$, shows agreement with the naive
  one-loop perturbative result.}
\label{kutishenfig}
\end{minipage}
\end{center}
\end{figure}

This task is made easier, and is also motivated, by the fact that
the standard one-doublet Higgs model {\it does not strictly 
exist} as a continuum field theory. This result is most easily
illustrated in terms of the Wilson renormalization group.\cite{wilson}
Any quantum field theory is defined using a regularization procedure
which ameliorates the bad short-distance behavior of the theory.
Following Wilson, we define the scalar sector of the standard
model
\beqa
{\cal L}_\Lambda =  & D^\mu \phi^\dagger D_\mu \phi + 
m^2(\Lambda)\phi^\dagger \phi 
+ {\lambda(\Lambda)\over 4}(\phi^\dagger\phi)^2 \\
& + {\eta(\Lambda)\over 36\Lambda^2}(\phi^\dagger\phi)^3+\ldots  
\nonumber
\eeqa
in terms of a {fixed} UV-cutoff $\Lambda$. Here we have allowed for
the possibility of terms of (engineering) dimension greater than four.
While there are an infinite number of such terms, one representative
term of this sort, $(\phi^\dagger \phi)^3$, has been included
explicitly for the purposes of illustration. Note that the coefficient
of the higher dimension terms includes the appropriate number of
powers of $\Lambda$, the intrinsic scale at which the theory is
defined.

Wilson observed that, for the purposes of describing experiments
at some fixed low-energy scale $E \ll \Lambda$, it is be possible to trade
a high-energy cutoff $\Lambda$ for one that is slightly
lower, $\Lambda^\prime$, so long as $E \ll \Lambda^\prime < \Lambda$.
In order to keep low-energy measurements fixed, it will in
general be necessary to redefine the values of the coupling constants
that appear in the Lagrangian. Formally, this process
is referred to as ``integrating out'' the
(off-shell) intermediate states with $\Lambda^\prime < k < \Lambda$. Keeping
the low-energy properties fixed we find
\beqa
{\cal L}_\Lambda & \Rightarrow & {\cal L}_{\Lambda^\prime} \nonumber\\
m^2(\Lambda)& \rightarrow & m^2(\Lambda^\prime) \\
\lambda(\Lambda) & \rightarrow & \lambda(\Lambda^\prime) \nonumber \\
\eta(\Lambda) & \rightarrow & \eta(\Lambda^\prime)  ~.
\nonumber
\eeqa

Wilson's insight was to see that many properties of the theory can be
summarized in terms of the evolution of these (generalized) couplings
as we move to lower energies. Truncating the infinite-dimensional
coupling constant space to the three couplings shown above, the
behavior of the scalar sector of the standard model is illustrated in
Figure \ref{3drg}. This figure illustrates a number of important
features of scalar field theory. As we flow to the infrared, {\it
  i.e.}  lower the effective cutoff, we find:
\begin{itemize}
\item $\eta \to 0$ --- this is the modern interpretation of
renormalizability. If $m_H \ll \Lambda$, the theory is drawn to
the two-dimensional $(m_H, \lambda)$ subspace. Any theory, therefore,
in which $m_H \ll \Lambda$ is close to a renormalizable theory with
corrections suppressed by powers of $\Lambda$.
\item $m^2 \to \infty$ --- This is the naturalness/hierarchy problem.
To maintain $m_H \simeq {\cal O}(v)$ we must adjust\,\footnote{Nothing we
discuss here will address the hierarchy problem directly.} the value
of $m_H$ in the underlying theory to of order
\beq
{\Delta m^2(\Lambda) \over m^2(\Lambda)} \propto {v^2 \over \Lambda^2}~.
\eeq
\item $\lambda \to 0$ --- The coupling $\lambda$ has a positive
  $\beta$ function and, therefore, as we scale to low energies
  $\lambda$ tends to 0.  If we try to take the ``continuum'' limit,
  $\Lambda \to +\infty$, the theory becomes free or
  trivial.\cite{wilson}
\end{itemize}

The triviality of the scalar sector of the standard one-doublet Higgs
model implies that this theory is only an effective low-energy theory
valid below some cut-off scale $\Lambda$.  Given a value of {$m^2_H =
  2 \lambda(m_H) v^2$}, there is an {\it upper} bound on {$\Lambda$}.
An {\it estimate} of this bound can be obtained by integrating the {one-loop} $\beta$-function, which yields
\beq
\Lambda \laem m_H \exp\left({4\pi^2v^2\over 3m^2_H}\right)~.
\label{landau}
\eeq
For a light Higgs, the bound above is at uninterestingly high scales
and the effects of the underlying dynamics can be too small to be
phenomenologically relevant. For a Higgs mass of order a few hundred
GeV, however, effects from the underlying physics can become
important. I will refer to these theories generically as ``composite
Higgs'' models.

Finally, while the estimate above is based on a perturbative analysis,
nonperturbative investigations of $\lambda \phi^4$ theory on the
lattice show the same behavior. This is illustrated in Figure
\ref{kutishenfig}.

\section{$T$, $S$, and $U$ in Composite Higgs Models}

In an $SU(2)_W \times U(1)_Y$ invariant scalar theory of a single doublet, all
interactions of dimension less than or equal to four also respect a
larger ``custodial'' symmetry\,\cite{custodial} which insures the
tree-level relation $\rho=M^2_W / M^2_Z \cos^2\theta_W\equiv 1$. The leading
custodial-symmetry violating operator is of dimension
six\,\cite{wyler} and involves four Higgs doublet fields
$\phi$. In general, the underlying theory does not respect the larger custodial
symmetry, and we expect the interaction
\beq
{\lower35pt\hbox{\epsfysize=1.00 truein \epsfbox{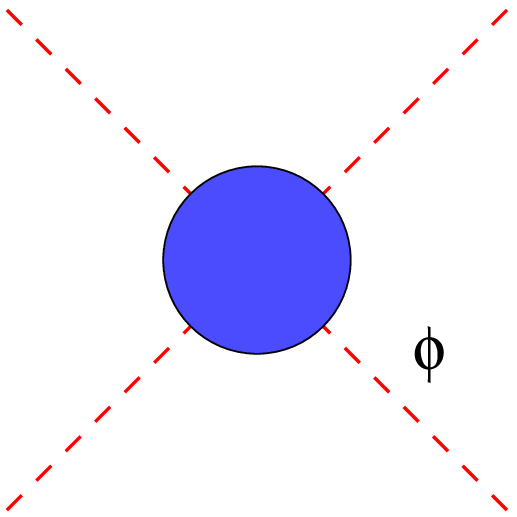}}}
\Rightarrow
{b { \kappa}^2 \over 2!\, { \Lambda}^2} 
(\phi^\dagger \stackrel{\leftrightarrow}{D^\mu} \phi)^2~,
\label{toperator}
\eeq
to appear in the low-energy effective theory.  Here b is an unknown
coefficient of ${\cal O}(1)$, and ${\kappa}$ measures size of
couplings of the composite Higgs field. In a strongly-interacting
theory, $\kappa$ is expected\,\cite{nda} to be of ${\cal O}(4\pi)$.

\begin{figure}[htb]
\begin{center}
\begin{minipage}[t]{7.5cm}
\includegraphics[width=7.5cm]{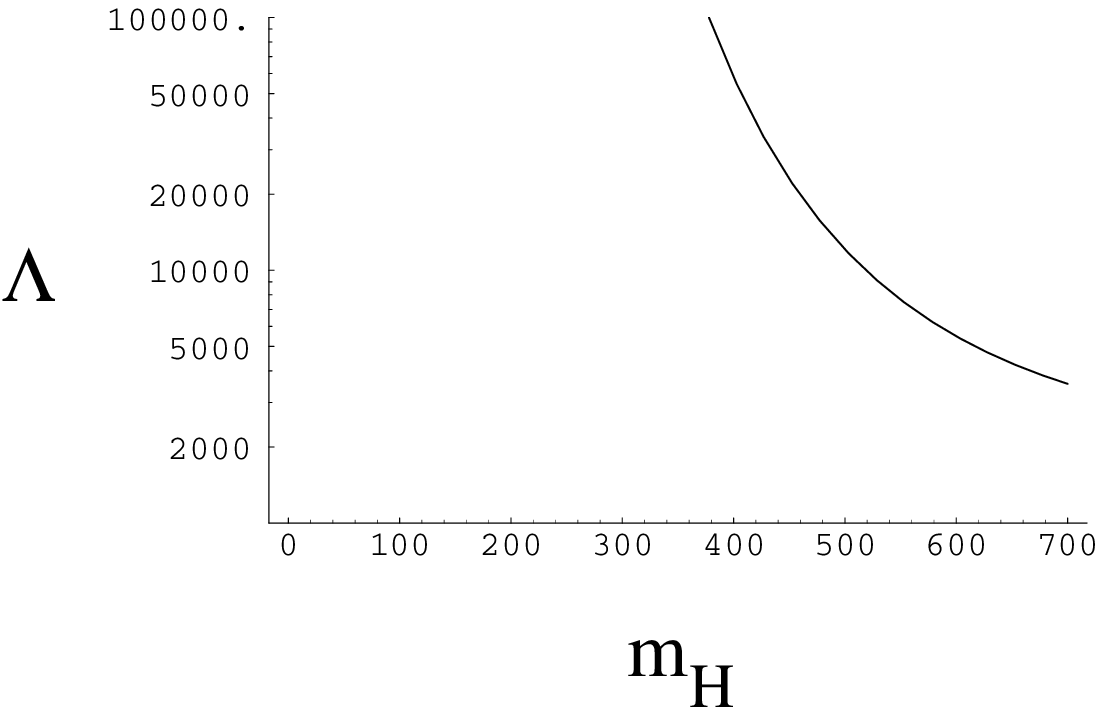}
\caption{Upper bound on scale $\Lambda$ as per eqn. (\protect\ref{landau}).}
\label{landaugraph}
\end{minipage}
\hspace{2mm}
\begin{minipage}[t]{7.5cm}
\includegraphics[width=7.5cm]{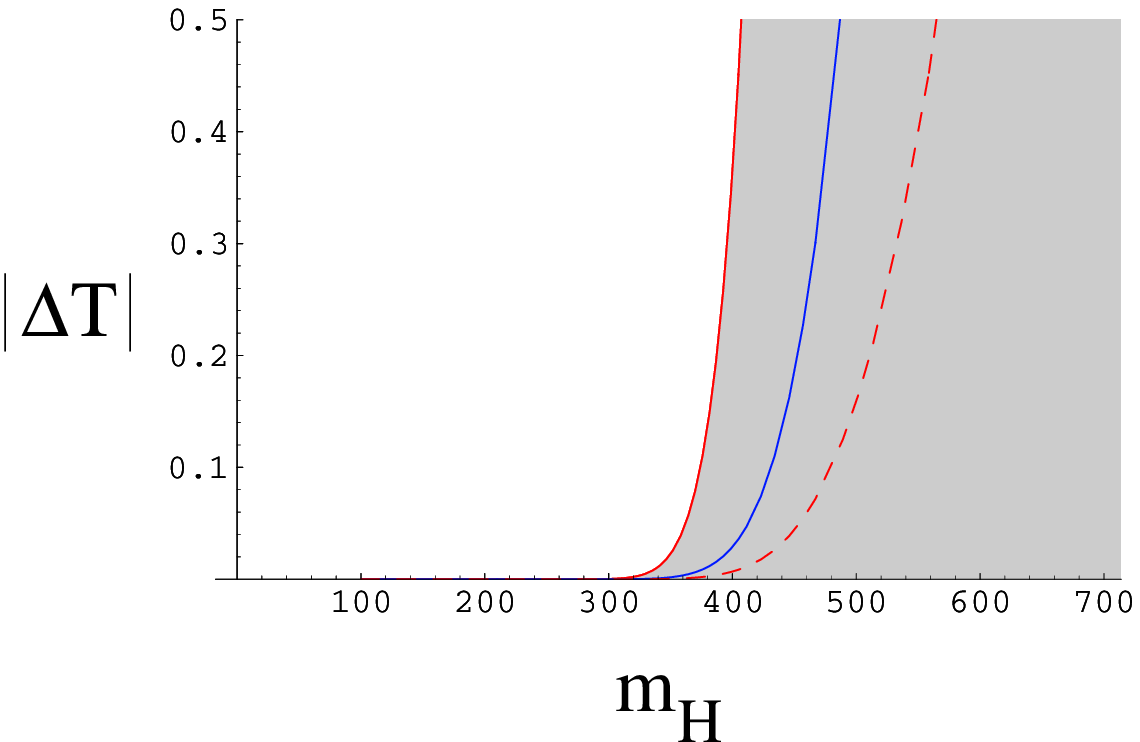}
\caption{Lower bound on expected size of $|\Delta T|$ as per eqn. (\protect\ref{tbound}),
for $|b|\kappa^2=16\pi^2$, $4\pi$, and 3.}
\label{tboundgraph}
\end{minipage}
\end{center}
\end{figure}

Deviations in the low-energy theory from the standard model can be
summarized in terms of the ``oblique'' parameters\,\cite{peskin,others}
$S$, $T$, and $U$.  The operator in eqn. \ref{toperator} will give
rise to a deviation ($\Delta \rho= \varepsilon_1 = \alpha T$)
\beq { |\Delta T|} = { |b| {\kappa}^2 {v^2 \over \alpha(M_Z)
    { \Lambda}^2}} { \gaem} {|b|\kappa^2\, v^2 \over \alpha(M^2_Z)\, m^2_H}\,
\exp\left({-\,{8 \pi^2 v^2\over 3 m^2_H}}\right) ~, 
\label{tbound}
\eeq
where $v \approx 246$ GeV and we have used eqn. \ref{landau} to obtain
the final inequality.  The consequences of eqns. (\ref{landau}) and
(\ref{tbound}) are summarized in Figures \ref{landaugraph} and
\ref{tboundgraph}. The larger $m_H$, the lower $\Lambda$ and the
larger the expected value of $\Delta T$.  Current limits imply $|T|
\stackrel{<}{\sim} 0.5$, and hence ${ \Lambda \stackrel{>}{\sim} 4\,
  {\rm TeV} \cdot \kappa}$.  (For $\kappa \simeq 4\pi$, {$m_H \laem
  450$ GeV}.)

By contrast, the leading contribution to $S$ arises from
\beq
{\lower35pt\hbox{\epsfysize=1.00 truein \epsfbox{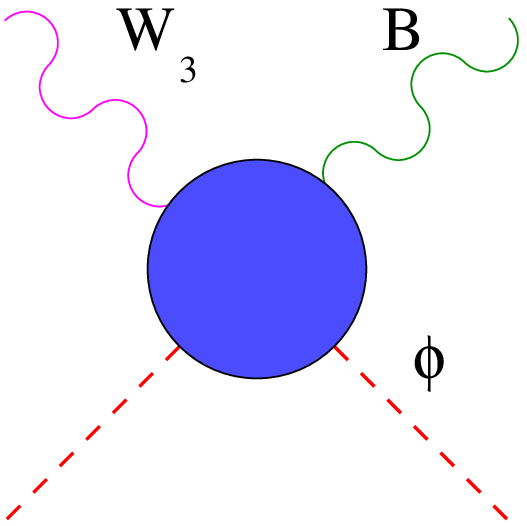}}}
\Rightarrow
-\,{a \over 2!\,{ \Lambda}^2} \left\{ [D_\mu,D_\nu] \phi \right\}^\dagger[D^\mu,D^\nu]\phi~.
\eeq
This gives rise to ($\varepsilon_3={\alpha S/4\sin^2\theta_W}$)
\beq \Delta S = {4 \pi a  v^2\over  { \Lambda}^2}~. \eeq
It is important to note that the size of contributions to $\Delta T$
and $\Delta S$ are very different
\beq 
{\Delta S \over \Delta T} = {a \over b} \left({4\pi \alpha \over { \kappa}^2}\right) ={\cal
  O}\left({10^{-1}\over { \kappa}^2}\right)~.
\eeq
Even for ${\kappa}\simeq 1$, $|\Delta S| \ll |\Delta T|$.

Finally, contributions to $U$ ($\varepsilon_2=-{\alpha U\over 4\sin^2\theta_W}$),
arise from
\beq
{c g^2 {\kappa^2}\over \Lambda^4} (\phi^\dagger W^{\mu\nu}\phi)^2
\eeq
and, being suppressed by $\Lambda^4$,  are typically much smaller than $\Delta T$.

\section{Limits on a Composite Higgs Boson}

\begin{figure}[htb]
\begin{center}
\begin{minipage}[t]{7.5cm}
\includegraphics[width=7.5cm]{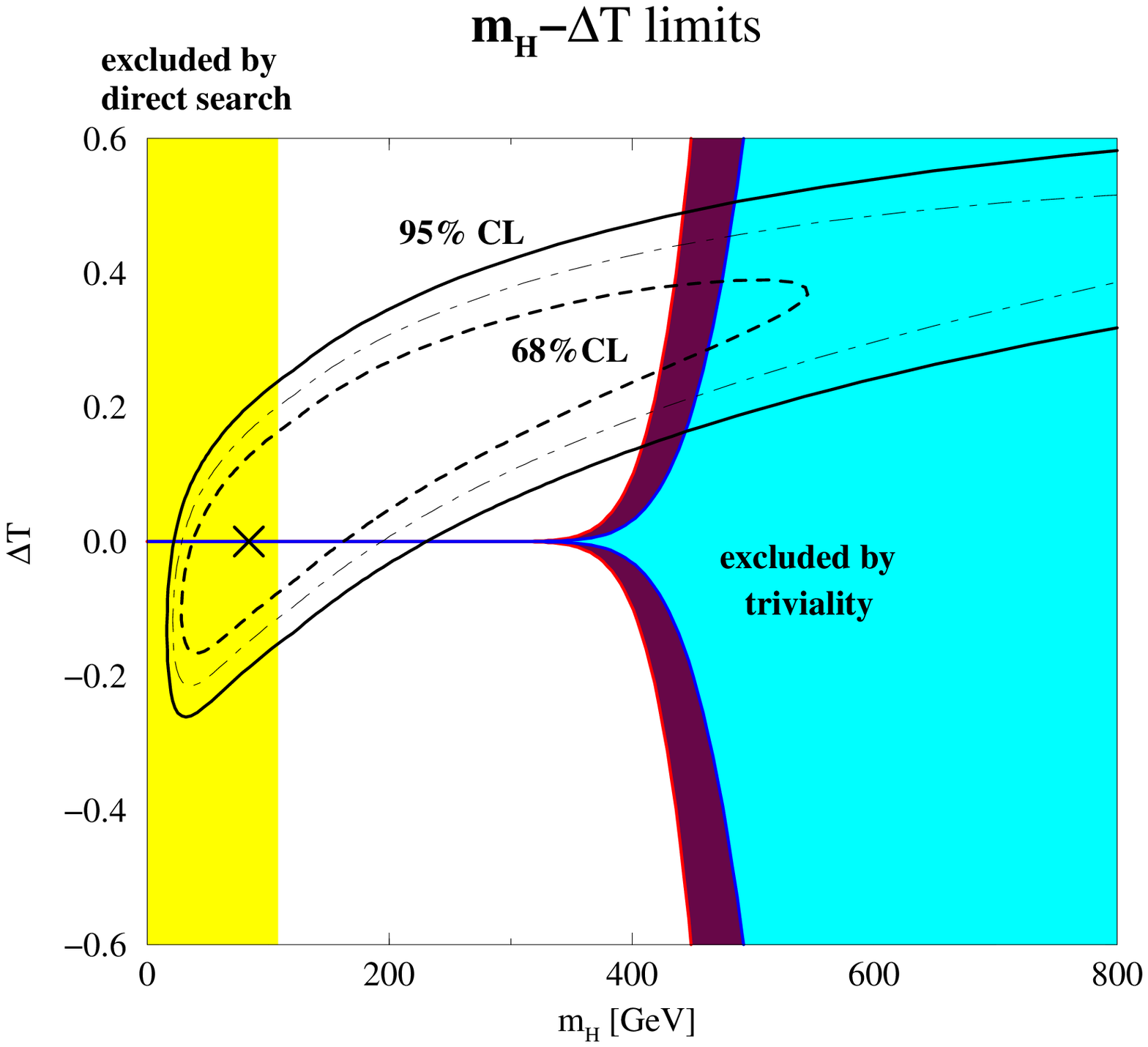}
\caption{68\% and 95\% CL regions allowed\,\protect\cite{prl} in $(m_H,|\Delta T|)$
  plane by precision electroweak
  data.\protect\cite{moriondewwg,lepewwg} Fit allows for $m_t$,
  $\alpha_s$, and $\alpha_{em}$ to vary consistent with current
  limits.\protect\cite{prl} Also shown by the dot-dash curve is the
  contour corresponding to $\Delta \chi^2=4$, whose intersection with
  the line $\Delta T=0$ -- at approximately 190 GeV -- corresponds to
  the usual 95\% CL upper bound quoted on the Higgs boson mass in the
  standard model. The triviality bound curves are for
  $|b|\kappa^2=4\pi$ and $4\pi^2$, corresponding to representative
  models.\protect\cite{prl}}
\label{mht_zfitter}
\end{minipage}
\hspace{2mm}
\begin{minipage}[t]{7.5cm}
\includegraphics[width=6.5cm]{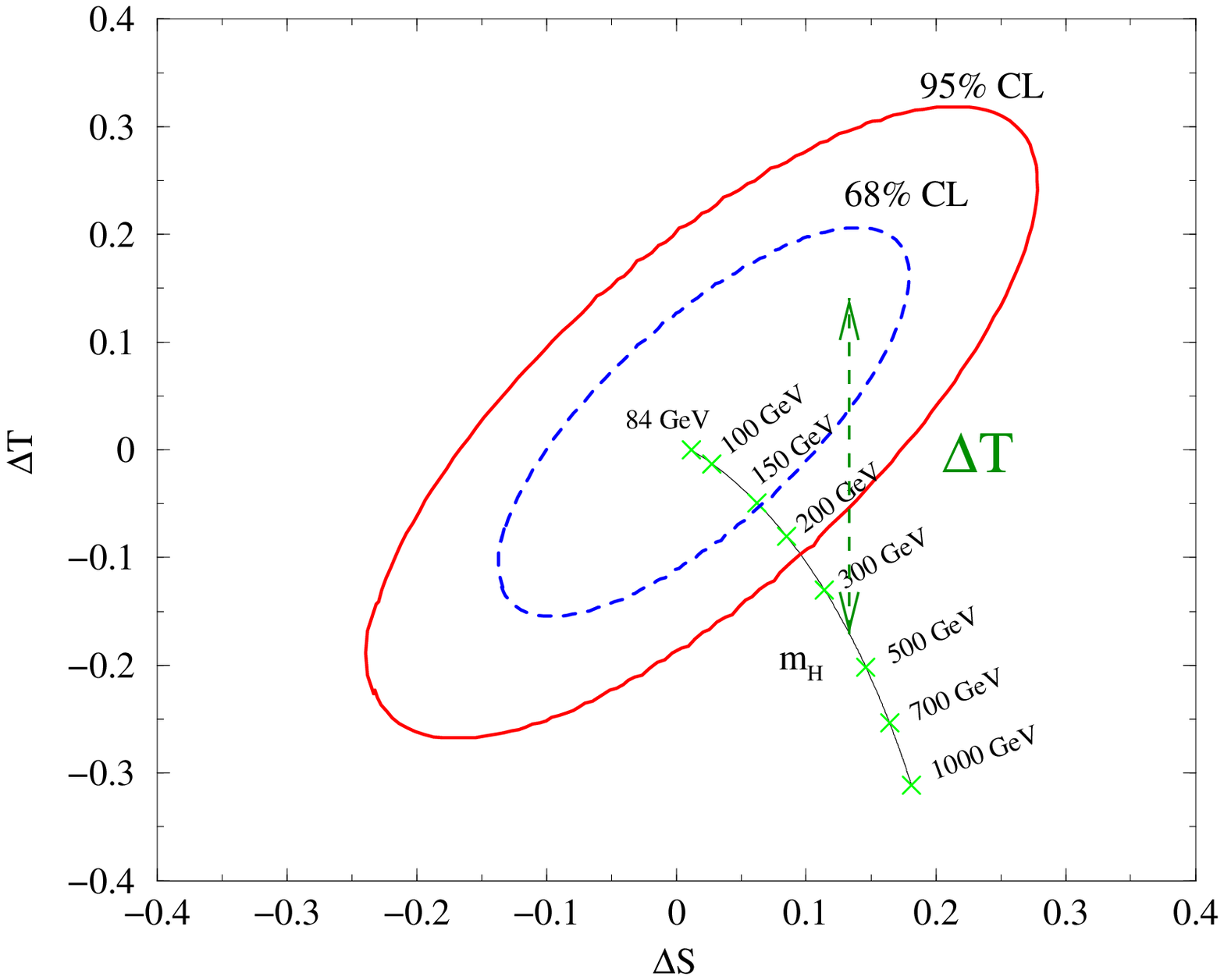}
\caption{68\% and 95\% CL regions allowed in $(\Delta S, \Delta T)$
plane by precision electroweak data.\protect\cite{moriondewwg,lepewwg}
Fit allows for $m_t$, $\alpha_s$, and $\alpha_{em}$ to vary consistent
with current limits.\protect\cite{prl} Standard model prediction for
varying Higgs boson mass shown as parametric curve, with $m_H$ varying from
84 to 1000 GeV.}
\label{stplot_varymt}
\end{minipage}
\end{center}
\end{figure}

From triviality, we see that the Higgs model can only be an effective
theory valid below some high-energy scale $\Lambda$.  As the Higgs
becomes heavier, the scale $\Lambda$ {\it decreases}. Hence, the
expected size of contributions to $T$ {\it grow}, and are larger than
the expected contribution to $S$ or $U$. The limits from precision
electroweak data in $(m_H, \Delta T)$ plane shown in Figure
\ref{mht_zfitter}.  We see that, for positive $\Delta T$ at 95\% CL,
the allowed values of Higgs mass extend to well beyond 800 GeV. On the
other hand, not all values can be realized consistent with the bound
given in eqn.  (\ref{landau}). As shown in figure \ref{mht_zfitter},
values of Higgs mass beyond approximately 500 GeV would likely require
values of $\Delta T$ much larger than allowed by current measurements.

I should emphasize that these estimates are based on dimensional
arguments, and we am  not arguing that it is {\it impossible} to
construct a composite Higgs model consistent with precision
electroweak tests with $m_H$ greater than 500 GeV.  Rather, barring
accidental cancellations in a theory without a custodial symmetry,
contributions to $\Delta T$ consistent with eqn. \ref{landau} are
generally to be expected. Specific composite Higgs boson models
are discussed in ref. 1, and the estimates given here are shown
to apply.

These results may also be understood by considering limits in the
$(S,T)$ plane for {\it fixed} $(m_H,m_t)$. In Figure
\ref{stplot_varymt}, changes from the nominal standard model best fit
($m_H=84$ GeV) value of the Higgs mass are displayed as contributions
to $\Delta S(m_H)$ and $\Delta T(m_H)$. Also shown are the 68\% and
95\% CL bounds on $\Delta S$ and $\Delta T$ consistent with current
data. We see that, for $m_H$ greater than ${\cal O}$(200 GeV), a
positive contribution to $T$ can bring the model within the allowed
region.

\begin{figure}[htb]
\begin{center}
\includegraphics[width=8cm]{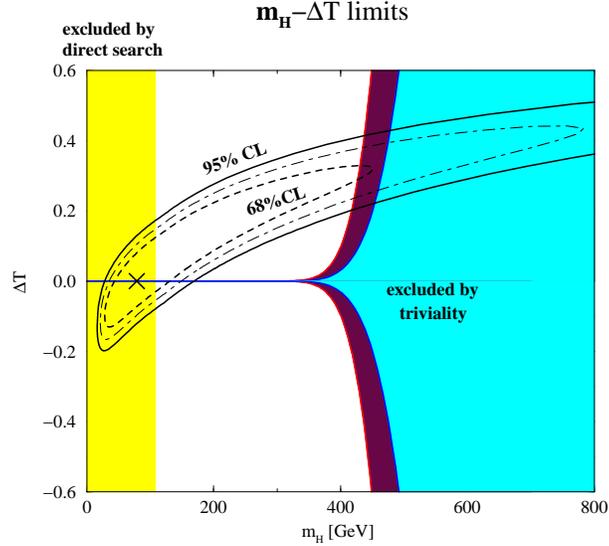}
\caption{68\% and 95\% CL allowed region\,\protect\cite{prl} in $(m_H,\Delta T)$ plane
  if uncertainty in top-quark and W-boson mass reduce to $\Delta m_t =
  2$ GeV and $\Delta M_W = 30$ MeV, as is possible at Run II of the
  Fermilab Tevatron.}
\label{mht_future}
\end{center}
\end{figure}

At Run II of the Fermilab Tevatron, it may be possible to reduce the
uncertainties in the top-quark and W-boson masses to $\Delta m_t = 2$
GeV and $\Delta M_W = 30$ MeV.\cite{runii}  Assuming that the measured values of
$m_t$ and $M_W$ equal their current central values, such a reduction
in uncertainties will result the limits in the $(m_H, \Delta T)$ plane
shown in Figure \ref{mht_future}. Note that, despite reduced
uncertainties, a Higgs mass of up to 500 GeV or so will still
be allowed.

\section{Conclusions}

In conclusion, the triviality of the Standard Higgs model implies that
it is at best a {low-energy effective} theory valid below a scale
{$\Lambda$} characteristic of nontrivial underlying dynamics.  As the
Higgs mass {increases}, the upper bound on the scale $\Lambda$
{decreases}. If the underlying dynamics does not respect a custodial
symmetry, it will give rise to {corrections to $T$ of order $\kappa^2
  v^2/\alpha \Lambda^2$}, while the contributions to $S$ and $U$ are
likely to be much smaller. For this reason, it is necessary to
consider limits on a Higgs boson in the {$(m_H,\Delta T)$} plane. In
doing so, we see that a Higgs mass {larger than 200 GeV is
  consistent} with precision electroweak tests {if} there is a
positive $\Delta T$. Absent a custodial symmetry, however, Higgs
masses {larger than $\simeq 500$ GeV are unlikely}: the scale of
underlying physics is so low that $\Delta T$ is likely to be too
large.

\section{Acknowledgments}

I thank Bogdan Dobrescu, Nick Evans, Christian H\"olbling, and
E.~H.~Simmons for fruitful collaborations. This work was supported in
part by the U.S. Department of Energy under grant DE-FG02-91ER40676.

\end{document}